# Gate-tunable Strong Spin-orbit Interaction in Two-dimensional Tellurium Probed by Weak-antilocalization

Chang Niu[1,2], Gang Qiu[1,2], Yixiu Wang[3], Zhuocheng Zhang[1,2], Mengwei Si[1,2], Wenzhuo Wu[3] and Peide D. Ye[1,2]

[1]School of Electrical and Computer Engineering, Purdue University, West Lafayette, Indiana 47907, USA

[2]Birck Nanotechnology Center, Purdue University, West Lafayette, Indiana 47907, USA

[3]School of Industrial Engineering, Purdue University, West Lafayette, Indiana 47907, USA

e-mail: Correspondence and requests for materials should be addressed P. D. Y. (yep@purdue.edu)

Abstract

Tellurium (Te) has attracted great research interest due to its unique crystal structure since 1970s. However, the conduction band of Te is rarely studied experimentally because of the intrinsic p-type nature of Te crystal. By atomic layer deposited dielectric doping technique, we are able to access the conduction band transport properties of Te in a controlled fashion. In this paper, we report on a systematic study of weak-antilocalization (WAL) effect in n-type two-dimensional (2D) Te films. We find that the WAL agrees well with Iordanskii, Lyanda-Geller, and Pikus (ILP) theory. The gate and temperature dependent WAL reveals that D'yakonov-Perel (DP) mechanism is dominant for spin relaxation and phase relaxation is governed by electron-electron (e-e) interaction. Large phase coherence length near 600nm at T=1K is obtained, together with gate tunable spin-orbit interaction (SOI). Transition from weak-localization (WL) to weak-antilocalization (WAL) depending on gate bias is also observed. These results demonstrate that newly developed solution-based synthesized Te films provide a new controllable strong SOI 2D semiconductor with high potential for spintronic applications.

Spin-orbit interaction (SOI) in two-dimensional (2D) materials brings many exotic phenomena to be explored. In transition-metal dichalcogenides (TMDs), large SOI induced band splitting in both conduction and valence band gives rise to valley Hall effect[1,2] and unconventional quantum Hall effect[3,4]. Recently, band inversion caused by spin-orbit coupling proximity effect[5] is observed in graphene/WSe$_2$ hetero-structure[6]. SOI has been extensively studied in III-V semiconductors like InGaAs/InAlAs quantum wells for spintronic applications[7,8]. Chiral crystals with SOI are predicted to host Kramers-Weyl fermions and other topological quantum properties[9].

Weak-antilocalization (WAL) and weak Localization (WL) caused by the interference of two time-reversal electron wave paths when electrons are scattered by impurities are used to probe SOI in conventional semiconductors[10] and now to be extended to 2D materials research such as graphene[11,12], MoS$_2$[13,14], black phosphorus[15–18], and others[19–22]. A correction to magneto-conductance due to the backscattered constructive or destructive interference between electrons is sensitive to the phase coherence and spin relaxation of electrons. WAL is also found in topologically nontrivial systems like topological insulators[23,24], Dirac[25] and Weyl[26] semimetals due to the significant Berry's phase.

In this paper, we perform magneto-transport measurements of 2D n-type Te at cryogenic temperatures. WAL is observed at low magnetic field less than 0.2T. Temperature and gate dependent WAL are systematically measured and analyzed. The spin-relaxation and phase-relaxation mechanisms are studied for the first time, showing the high-quality of the 2D Te film and the existing strong SOI in this material.

Te is a narrow bandgap (0.35eV) semiconductor with hexagonal crystal structure formed by van der Waals interaction between each one-dimensional helical atom chain (Figure 1a). Covalently bonded atoms rotate around c axis in a period of three atoms as shown in Figure 1a. The valence band and conduction band are located in the corner of Brillouin zone H and H' points (Figure 1b). Theoretically, Te is predicted to undergo transformation into a topological insulator under strain[27] and Weyl semimetal under pressure[28]. Te is intrinsically p-type doped, therefore up to date most of the experiments including thermoelectric properties[29], quantum Hall effect[30] and angle-resolved photoemission spectroscopy (ARPES)[31] were performed in p-type Te samples. The lack of inversion symmetry and the strong spin-orbit coupling of Te give rise to the camelback-like structure in valence band[30] and Rashba-like spin splitting bands with nontrivial radial spin texture in conduction band[28,32]. The spin-split conduction bands cross at H point and form a Weyl point protected by the three-fold screw symmetry of the helical crystal[27]. But the spin-orbit interaction and its mechanism in n-type Te remains unexplored.

The inset of Figure 1d is an optical image of Te device with Hall-bar structure for magneto-transport studies. Source and drain electrodes were made along the long edge of the flake which coincides with the direction of the atomic chains[33]. Hall bars are used to measure the longitudinal and transverse resistance with the back-gate to tune the film electron density. The sketch of typical n-type field-effect device is shown in Figure 1c. 20nm $Al_2O_3$ is grown by atomic layer deposition (ALD) at 200 °C to dope tellurium into n-type[34,35]. Similar effect has also been observed in black phosphorus[36,37]. In order to prove the n-type doping effect, a transfer curve of a Te field-effect device is measured (Figure 1d) using four-terminal method. By carrying out the Hall measurement, we can calculate the two-dimensional carrier density $n$ and electron Hall mobility $\mu$ under different gate biases (Figure 1e) using $n = \frac{1}{e}\left(\frac{dB}{dR_{xy}}\right)$ and $\mu =$ $\left(\frac{L}{W}\right)\left(\frac{1}{R_{xx}ne}\right)$, where $L$ is channel length, $W$ is channel width, $R_{xx}$ is longitudinal resistance, and $R_{xy}$ is the Hall resistance. More than 10 devices were fabricated and measured. All the results presented below is from one representative device. Another n-type Te WAL device is analyzed in Supporting Note 3. Other devices have reproducible phenomena and similar results.

Two different spin-relaxation mechanisms can be used to explain the formation of WAL: the D'yakonov-Perel (DP)[38] and the Elliot-Yafet (EY) spin-relaxation mechanisms. EY spin-relaxation mechanism often exists in spin degenerated bands in which spin-up and spin-down states are entangled by SOI. When an elastic scattering (momentum scattering) process occurs, the electron spin may flip during the scattering. The DP spin-relaxation is attributed to the spin precession process caused by an effective magnetic field $\overrightarrow{B_{eff}}$ ($\overrightarrow{B_{eff}} \propto \vec{E} \times \vec{p}$ [39,40] where $\vec{E}$ is the electric field and $\vec{p}$ is the electron momentum) which is induced by Rashba[41] and/or Dresselhaus SOI[42]. The electron spin makes the precession along the effective magnetic field direction during elastic scattering process. The scattering of an electron by impurities and phonons which changes the electron momentum $\vec{p}$, fluctuate the effective magnetic field $\overrightarrow{B_{eff}}$ and suppress the spin relaxation. Therefore, the spin-relaxation time $\tau_{so}$ is proportional to momentum scattering time $\tau_{tr}$ for EY spin-relaxation mechanism and inversely proportional to $\tau_{tr}$ for DP mechanism[20,43]. Rashba SOI induced by the external electric field and Dresselhaus SOI induced by crystal electric field in inversion asymmetric materials contribute to band spin-splitting even at zero external magnetic field.

WAL based on EY spin-relaxation mechanism is described by Hikami, Larkin, and Nagaoka (HLN) theory[44], meanwhile, Iordanskii, Lyanda-Geller, and Pikus (ILP)[45] theory is based on DP mechanism. Both theories are valid when the external magnetic field $B$ is smaller than the characteristic magnetic field for transport $B_{tr}$ ($B_{tr} = \hbar/4eD\tau_{tr}$ where $D$ is the diffusion constant and $\tau_{tr}$ is the momentum scattering time). $B_{tr}$ in n-type Te is determined to be 0.2T (see Supporting Note 2). All the results in this work are from the fitting of data below 0.2T. Fittings of Te WAL experiment data with both theories are shown in Figure 2a, where red line presents for ILP theory and blue line for HLN theory. It is clear that ILP fitting provides better agreement with experiment data, indicating that DP spin-relaxation mechanism is dominant in Te single crystal films.

Magneto conductance of WAL using ILP theory is expressed as shown below[38,43]:

$$\Delta\sigma_{xx}(B) - \Delta\sigma_{xx}(0)$$
$$= \frac{e^2}{2\pi^2\hbar}\left\{\Psi\left(\frac{1}{2} + \frac{B_\varphi}{B} + \frac{B_{so}}{B}\right) - \ln\frac{B_\varphi + B_{so}}{B} + \frac{1}{2}\Psi\left(\frac{1}{2} + \frac{B_\varphi}{B} + \frac{2B_{so}}{B}\right)\right.$$
$$\left. - \frac{1}{2}\ln\frac{B_\varphi + 2B_{so}}{B} - \frac{1}{2}\Psi\left(\frac{1}{2} + \frac{B_\varphi}{B}\right) + \frac{1}{2}\ln\frac{B_\varphi}{B}\right\}$$

$$B_x = \frac{\hbar}{4eD\tau_x} \quad L_x = \sqrt{D\tau_x} \quad x = so, \varphi \qquad (1)$$

Where $e$ is the elementary charge, $\hbar$ is the reduced Planck constant, $\Psi$ is the digamma-function, $D$ is the diffusion constant, $\tau_\varphi$ is the phase-relaxation time, $\tau_{so}$ is the spin-relaxation time, $L_{so}$ and $L_\varphi$ are spin-relaxation length and phase coherence length respectively. $B_\varphi$ and $B_{so}$ are the only two fitting parameters. From the ILP fitting (see Supporting Note 2) we conclude that $k$-linear SOI effect is much smaller than $k$-cubic SOI effect in our Te sample which is different from III-V semiconductors.

In order to extract $\tau_{so}$ from $B_{so}$, diffusion constant $D$ ($D = v_F^2\tau_{tr}/2$, where $v_F$ is the Fermi velocity) and effective mass $m^*$ are calculated from Hall measurement and temperature dependent Shubnikov-de Haas (SdH) oscillations (Supporting Note 1). We are able to extract the effective mass of electrons $m_e^* = 0.11m_0$, where $m_0$ is the bare electron mass (Figure S1). It is consistent with the previous theoretical prediction[46]. The Te conduction band minimum is at H point of the Brillouin zone (Figure 1b) which has two-fold valley degeneracy and two-fold spin degeneracy[28]. To better understand the spin-relaxation mechanism in Te, the relation between spin-relaxation time $\tau_{so}$ and momentum scattering rate $\tau_{tr}^{-1}$ is presented in Figure 2b. The red eye-guideline indicates that $\tau_{so}$ is inversely proportional to $\tau_{tr}$. This result together with the better fitting of experimental data with ILP theory unveils that DP spin-relaxation mechanism (spin-precession) is dominant in n-type 2D Te films.

Gate dependent WAL is measured at $T = 1K$ (Figure 3a). $B_\varphi$ $(L_\varphi)$ and $B_{so}$ $(L_{so})$ are extracted from WAL curves (Figure 3b,c). The phase coherence length $L_\varphi$ reaches as high as 573nm at $V_{bg} = 30V$ which is larger than that of p-type Te (Figure S6b)[47] and other 2D materials[13,14,19–21,40,47–50,51] listed in Table 1. It is worth to mention that the elastic scattering length is extracted to be $L_e$ (23-47nm) which is one-order of magnitude smaller than the phase coherence length in all gate voltage. Thus, the electron transport in n-type Te at low temperatures is in quantum diffusive regime. In addition, the decrease of spin-relaxation length $L_{so}$ with gate voltage can only be explained by the decrease in $\tau_{so}$ (stronger SOI) at higher gate voltage because D is increasing with gate voltage according to $L_{so} = \sqrt{D\tau_{so}}$ and $D = v_F^2 \tau_{tr}/2$. This further confirms that spin-splitting can be tuned by gate in n-type Te. We are not able to distinguish Rashba and Dresselhaus SOI, since both effects have k-cubic dependent term[43] and the strength is increasing with gate voltage thus carrier density. The gate-tunable SOI together with the long phase coherence length gives 2D Te a new edge for spintronic applications over other 2D materials that either lack of controllable SOI (like graphene[11,12]) or suffer from short phase coherence length(Table 1). The phase coherence length $L_\varphi$ increases with back gate-voltage V$_{bg}$, implying that electron-electron scattering is the main phase-relaxation mechanism. We will further discuss the dephasing mechanism below, which plays an important role in the formation of WAL.

In general, phase coherence length is limited by inelastic electron-electron scattering and electron-phonon (e-ph) scattering[19]. For inelastic electron-electron scattering with small energy transfer, phase coherence length $L_\varphi$ is expressed as[16]

$$L_\varphi = \frac{h^2 \sigma_{xx}}{\pi e^2}(m^* k_B T ln \frac{\sigma_{xx} h}{e^2})^{-1/2} \quad (2)$$

by the Altshuler-Aronov-Khmelnitsky (AAK) theory[52]. By plotting experimentally extracted $L_\varphi$ as a function of conductance $\sigma_{xx}$ (Figure 4c), we can confirm that the near linear dependence between gate voltage and phase coherence length at fixed temperature originates from electron-electron interaction. To further investigate the inelastic scattering mechanism of Te, temperature dependent WAL is also measured as shown in Figure 4a. When the temperature increases, WAL peaks attenuate because of the decrease in phase coherence length (Figure 4b). The temperature dependent phase coherence length $L_\varphi \propto T^{-\gamma}$ distinguishes different scattering mechanism. For example, $\gamma = 1$ is for e-ph interaction and $\gamma = 0.5$ for e-e interaction[19,21]. Figure 4b inset presents gate dependent $\gamma$ extracted from the power law fitting of $L_\varphi$. Measured values of $\gamma$ are close to 0.5 in all gate range which is in a good agreement with AAK theory described in Eq. 2. The slightly deviation from 0.5 can be explained by the temperature dependence of $\sigma_{xx}$. Therefore, we conclude that electron-electron interaction is the main phase-relaxation mechanism in Te within temperatures ranging from 1K to 18K.

Interestingly, quantum interference effects like WL and WAL are sensitive to the relative strength of spin-relaxation and phase coherence as we discussed before. System shows

WAL when the phase coherence length is larger than the spin-relaxation length and WL otherwise[40]. The gate dependent transition from WAL to WL is observed at higher temperature (18K) (Figure 4d). According to the AAK theory (Eq. 2), the phase coherence length $L_\varphi$ is small in the region of high temperatures and low gate voltage (low longitudinal conductance $\sigma_{xx}$). In the meantime, spin-relaxation length $L_{so}$ decreases with gate voltage (Figure 3c) because of the Rashba and Dresselhaus SOI. Therefore, WL is observed when the phase coherence length $L_\varphi < L_{so}$ at $V_{bg} = 14V$. With the increase of gate voltage, WAL emerges under the condition of $L_\varphi > L_{so}$ at $V_{bg} > 18V$. WAL is also observed in p-type 2D Te films as shown in Supporting Note 4.

In conclusion, gate-tunable strong spin-orbit interaction induced WAL has been observed in n-type 2D Te films. Gate and temperature dependent WAL are systematically measured and analyzed. We find that k-linear spin-precession vector is much smaller than k-cubic spin-precession vector by fitting experimental data with ILP theory. Large phase coherence length near 600nm together with gate-tunable SOI makes n-type 2D Te films a competitive candidate for spintronic applications. Furthermore, we determined that DP spin-relaxation mechanism is dominant and electron-electron interaction with small energy transfer is the main mechanism for inelastic scattering at low temperatures for 2D Te films.

**Table 1** Phase coherence length $L_\varphi$ in 2D materials

| Materials | $L_\varphi$(nm) | References |
|---|---|---|
| p-type Te | 63.6 | 47 |
| MoS$_2$ | 50 | 13,14 |
| Black phosphorus | 104 | 15-18 |
| InSe | 320 | 21 |
| GaSe | 130 | 20 |
| WSe$_2$ | 140 | 40 |
| VSe$_2$ | 50 | 48 |
| TaSe$_2$ | 100 | 22 |
| Nb$_3$SiTe$_6$ | 70 | 19 |
| Bi$_2$O$_2$Se | 300 | 51 |
| MoTe$_2$ | 80 | 49,50 |

Figures

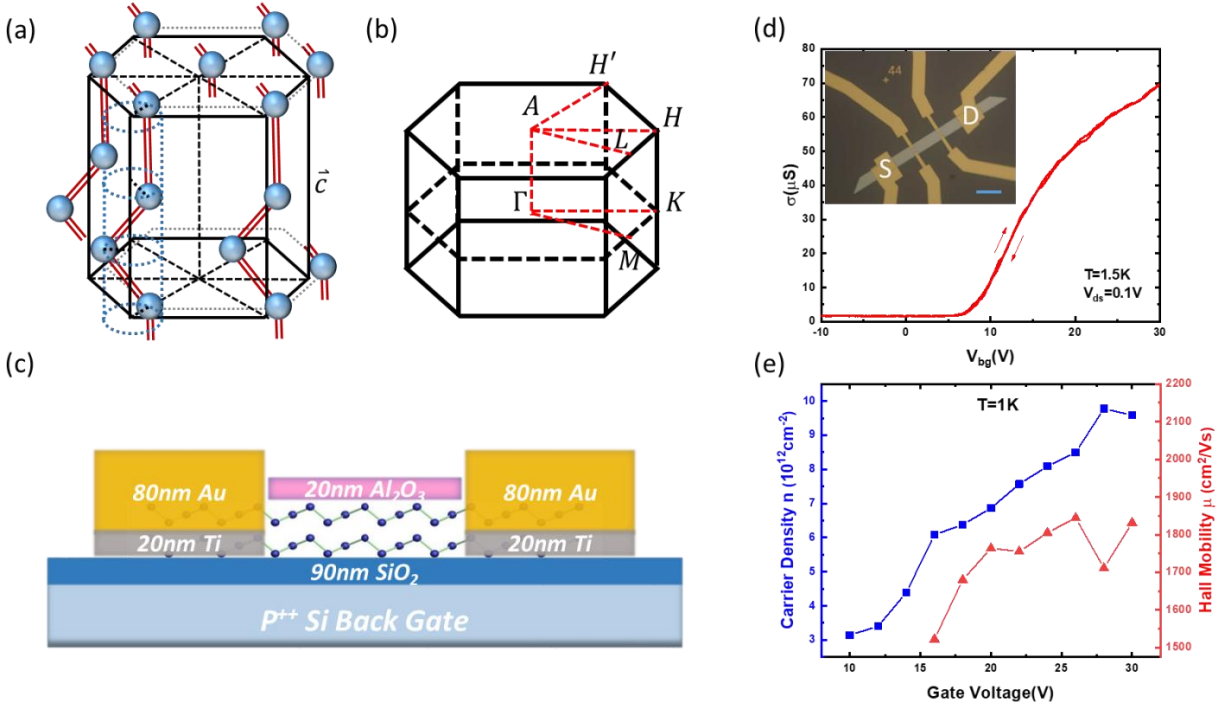

**Figure 1 (a)** Crystal structure of Tellurium (Te). Te has one-dimensional helical atom chains along $\vec{c}$ direction and hexagonal structure is formed by van der Waals interaction between each chain. **(b)** The first Brillouin zone of Te. The Te conduction band minima is located in H and H' points. **(c)** Device structure sketch of an n-type Te Field-effect transistor. **(d)** Transfer curve of n-type Te field-effect device with Te film thickness of 10nm at 1.5K. **Inset:** An optical image of Hall-bar device for transport measurement with helical atom chains (c-axis) along long edge of the flake. The scale bar is 20µm. **(e)** Gate-dependent carrier density $n$ (blue squares, left axis) and electron mobility $\mu$ (red triangles, right axis) extracted from Hall-measurement at low 1K.

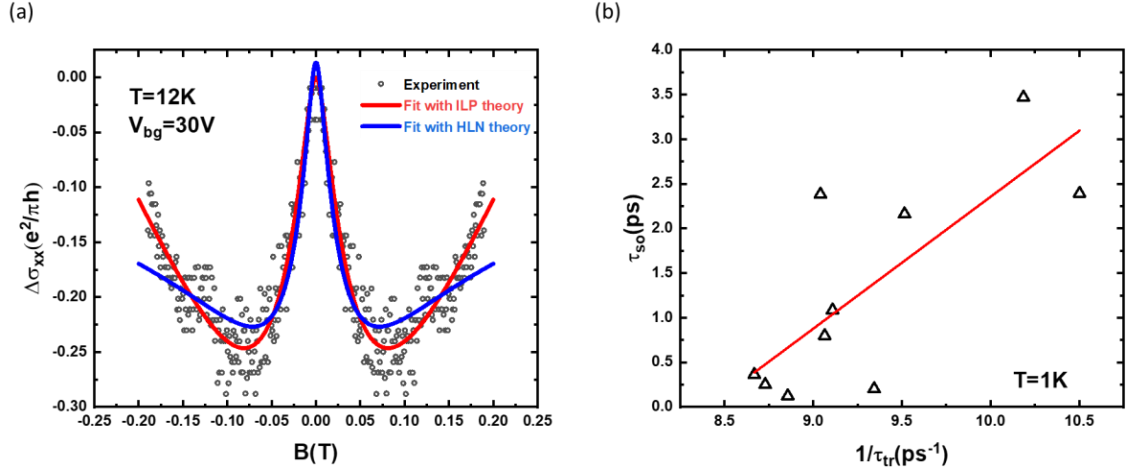

**Figure 2 (a)** Experimental Weak-antilocalization (WAL) data (black circles) measured at $T = 12K$ $V_{bg} = 30V$ and fitting curves with different theory (red line for ILP theory, blue line for HLN theory). IPL theory (red line) provides better agreement with experimental data. **(b)** Spin-relaxation time $\tau_{so}$ as a function of momentum scattering rate $\tau_{tr}^{-1}$ at $T = 1K$. The red solid line is the linear fit of data points. The positive correlation between $\tau_{so}$ and $\tau_{tr}^{-1}$ indicates the D'yakonov-Perel' (DP) spin-relaxation mechanism is dominant in Te.

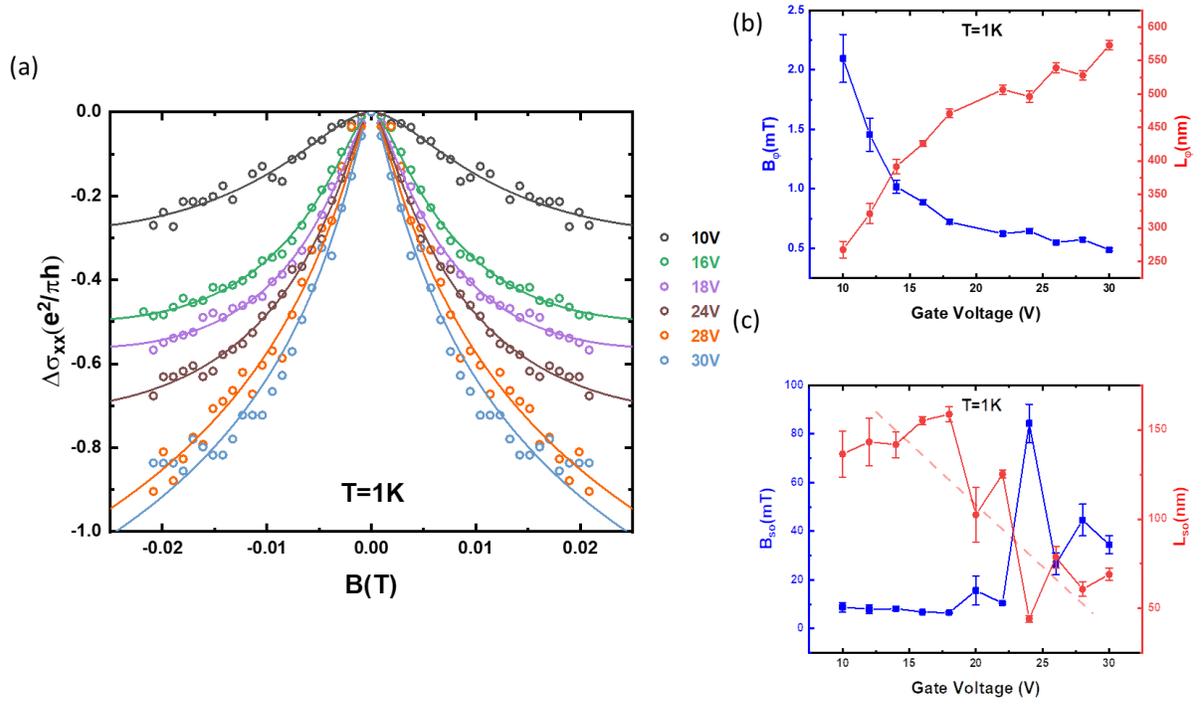

**Figure 3 Gate dependence of WAL. (a)** Theoretical fits (solid lines) with ILP theory to experimental data (circles) measured at various back-gate voltages. **(b)** Gate-dependent $B_\varphi$ (blue squares, left axis) and phase coherence length $L_\varphi$ (red squares, right axis) extracted from WAL fitting at $T = 1K$. **(c)** Gate-dependent $B_{so}$ (blue squares, left axis) and spin-relaxation length $L_{so}$ (red squares, right axis) extracted from WAL fitting at $T = 1K$.

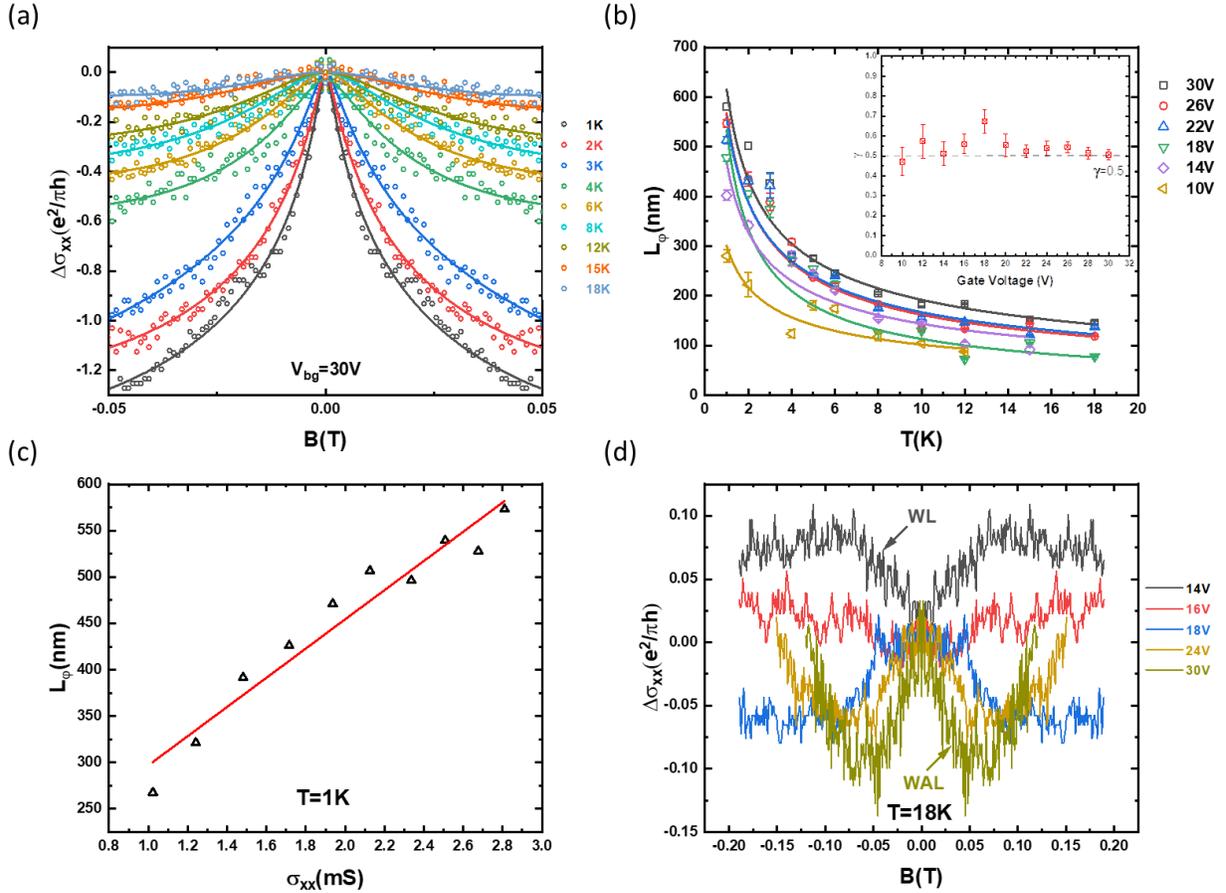

**Figure 4 Temperature dependence of WAL (a)** Theoretical fits (solid lines) with ILP theory to temperature dependent experimental data (circles) measured at $V_{bg} = 30V$. **(b)** Temperature dependent phase coherence length $L_\varphi$ (data points) extracted from WAL at different gate-voltage fit with power law (solid lines). **Inset**: Gate-dependent coefficient ϒ extracted from the power law fitting are close to 0.5 in all gate-voltage which indicates the electron-electron interaction is responsible for the dephasing process. **(c)** Phase coherence length $L_\varphi$ as a function of the sheet conductance $\sigma_{xx}$ at $T = 1K$. The red line is the linear fitting of $L_\varphi$ and $\sigma_{xx}$. **(d)** Transition from Weak-localization (WL) to Weak-antilocalization (WAL) at $T = 18K$.

Supporting Information for:

# Gate-tunable Strong Spin-orbit Interaction in Two-dimensional Tellurium Probed by Weak-antilocalization

Chang Niu[1,2], Gang Qiu[1,2], Yixiu Wang[3], Zhuocheng Zhang[1,2], Mengwei Si[1,2], Wenzhuo Wu[3] and Peide D. Ye[1,2]

[1]School of Electrical and Computer Engineering, Purdue University, West Lafayette, Indiana 47907, USA

[2]Birck Nanotechnology Center, Purdue University, West Lafayette, Indiana 47907, USA

[3]School of Industrial Engineering, Purdue University, West Lafayette, Indiana 47907, USA

e-mail: Correspondence and requests for materials should be addressed P. D. Y. (yep@purdue.edu)

# Supporting Note 1: Shubnikov-de Haas (SdH) oscillations under high magnetic field and effective mass of electrons

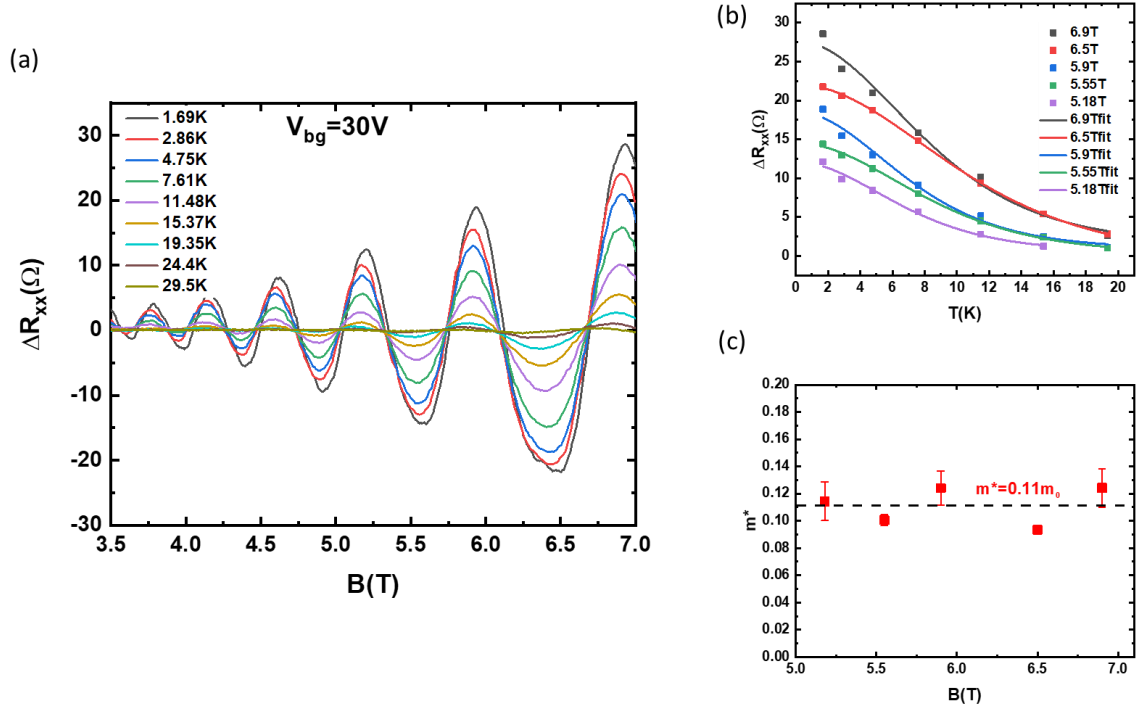

**Figure S1 (a)** Temperature dependent Shubnikov-de Haas (SdH) oscillation amplitudes of n-type Tellurium. **(b)** Extracted oscillation amplitudes under different magnetic field as a function of temperature (squares). The solid lines are fitting curves with Lifshitz-Kosevich[1] equation. **(c)** Effective cyclotron mass is $0.11m_0$ calculated from (b).

Because of the high mobility, SdH oscillations are observed at high magnetic field. We subtracted the background magneto resistance (MR) and obtained SdH oscillation amplitude as shown in Figure S1a. By fitting each temperature dependent SdH oscillation with Lifshitz-Kosevich equation[1] (Figure S1b)

$$\Delta R_{xx} \propto \frac{2\pi^2 m^* k_B T / \hbar e B}{\sinh(2\pi^2 m^* k_B T / \hbar e B)}$$

(where $k_B$ is the Boltzmann constant and $m^*$ is cyclotron effective mass)

**Supporting Note 2: Fitting of Iordanskii, Luanda-Geller, and Pikus (ILP) theory**

Iordanskii, Lyanda-Geller, and Pikus (ILP) theory[2–4]

$$\Delta\sigma_{xx}(B) - \Delta\sigma_{xx}(0) = -\frac{e^2}{4\pi^2\hbar}\left[\frac{1}{a_0} + \frac{2a_0+1+\frac{B_{so}}{B}}{a_1\left(a_0+\frac{B_{so}}{B}\right)-2\frac{B_{so1}}{B}} - \sum_{n=1}^{\infty}\left\{\frac{3}{n} - \frac{3a_n^2+2a_n\frac{B_{so}}{B}-1-2(2n+1)\frac{B_{so1}}{B}}{\left(a_n+\frac{B_{so}}{B}\right)a_{n-1}a_{n+1}-2\frac{B_{so1}}{B}[(2n+1)a_n-1]}\right\} + \Psi\left(\frac{1}{2}+\frac{B_\varphi}{B}\right) + 2\ln\frac{B_\varphi}{B} + 3C\right] \quad (S1)$$

where $a_n = n + \frac{1}{2} + \frac{B_\varphi}{B} + \frac{B_{so}}{B}$   $B_{so} = B_{so1} + B_{so3}$   $B_\varphi = \frac{\hbar}{4eD\tau_\varphi}$

$B_{so1} = \frac{\hbar}{4eD}|\Omega_1|^2\tau_1$   $B_{so3} = \frac{\hbar}{4eD}|\Omega_3|^2\tau_3$   $\frac{1}{\tau_m} = \int(1-cosm\theta)W(\theta)d\theta$ $(m=1,3)$  $\Psi$ is digamma-function, C is the Euler's constant, D is diffusion constant, $W(\theta)$ is the probability of scattering by angle $\theta$, $\Omega_1$ and $\Omega_3$ are the spin-precession vector originated from Rashba and Dresselhaus SOI where $\Omega_1$ is linear in $k$ and $\Omega_3$ is cubic in $k$. To calculate Eq. S1 the infinite sum is added up to $n = 2\times 10^5$. The theory is valid when magnetic field B is smaller than characteristic magnetic field $B_{tr} = \hbar/4eD\tau_{tr}$ which is calculated to be around 0.2T from Hall and SdH oscillation measurement.

Rashba and Dresselhaus SOI are expected in 2D-Te FETs because of the lack of inversion symmetry and the presence of strong electric field from the gate. Taking in to account the influence of both SOI, WAL is described by ILP theory (Eq. S1) with three parameters $B_\varphi, B_{so1}$ and $B_{so3}$ ($B_{som} = \frac{\hbar}{4eD}|\Omega_m|^2\tau_m$ $m = 1,3$ where $\Omega_1$ is the $k$-linear dependent spin-precession vector and $\Omega_3$ is the $k$-cubic dependent spin-precession vector, $\tau_1 = \tau_3 = \tau_{tr}$ if we assume isotropic scattering for simplicity.

When $B_{so1} = 0$ ($\Omega_1 = 0$), the equation reduced to:

$$\Delta\sigma_{xx}(B) - \Delta\sigma_{xx}(0) = \frac{e^2}{2\pi^2\hbar}\left\{\Psi\left(\frac{1}{2}+\frac{B_\varphi}{B}+\frac{B_{so}}{B}\right) - \ln\frac{B_\varphi+B_{so}}{B} + \frac{1}{2}\Psi\left(\frac{1}{2}+\frac{B_\varphi}{B}+\frac{2B_{so}}{B}\right) - \frac{1}{2}\ln\frac{B_\varphi+2B_{so}}{B} - \frac{1}{2}\Psi\left(\frac{1}{2}+\frac{B_\varphi}{B}\right) + \frac{1}{2}\ln\frac{B_\varphi}{B}\right\} \quad (S2)$$

We use the reduction form of ILP theory to fit the experimental data in the main text. First because the fitting of the WAL data with (Eq. S2) is good. Second in our system the k linear term is much smaller than k cubic term since the density is relatively high ($3\times 10^{12} cm^{-2}$) which means $k$ is large in our density range.

Here, we analyze our data with Eq. S1 including all three parameters ($B_\varphi$, $B_{so1}$ and $B_{so3}$). We define $r = B_{so1}/B_{so3}$ to describe the strength of two SOIs. First, several values of $r$ ($r = 0$ means only $k$ cubic term exist and $r = \infty$ means only $k$ linear term exist) are set to fit the experimental data (Figure S2). The fitting parameters are presented in table S1. Second, fitting curves of each $r$ with different $B_\varphi$ are shown in Figure S3. There results indicate that Rashba SOI ($k$ linear term) is small in our system.

**Table S1** Parameters used for the fitting in Figure S2

| $r = B_{so1}/B_{so3}$ | $B_\varphi$(mT) | $B_{so1}$(mT) | $B_{so3}$(mT) |
|---|---|---|---|
| 0 | 5.6 | 0 | 20.6 |
| 0.1 | 5.6 | 2 | 20.1 |
| 0.2 | 5.6 | 3.8 | 19.1 |
| 0.5 | 5.6 | 7.8 | 15.6 |
| 1 | 5.6 | 11.6 | 11.6 |
| ∞ | 5.6 | 11.5 | 0 |

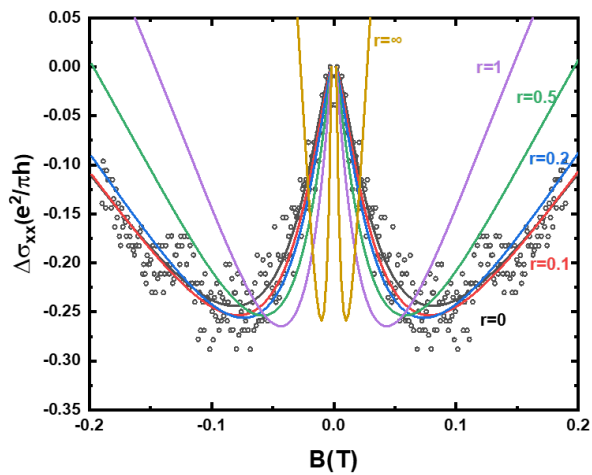

**Figure S2** Theoretical fits calculated from ILP theory Eq. S1 with different values of $r$

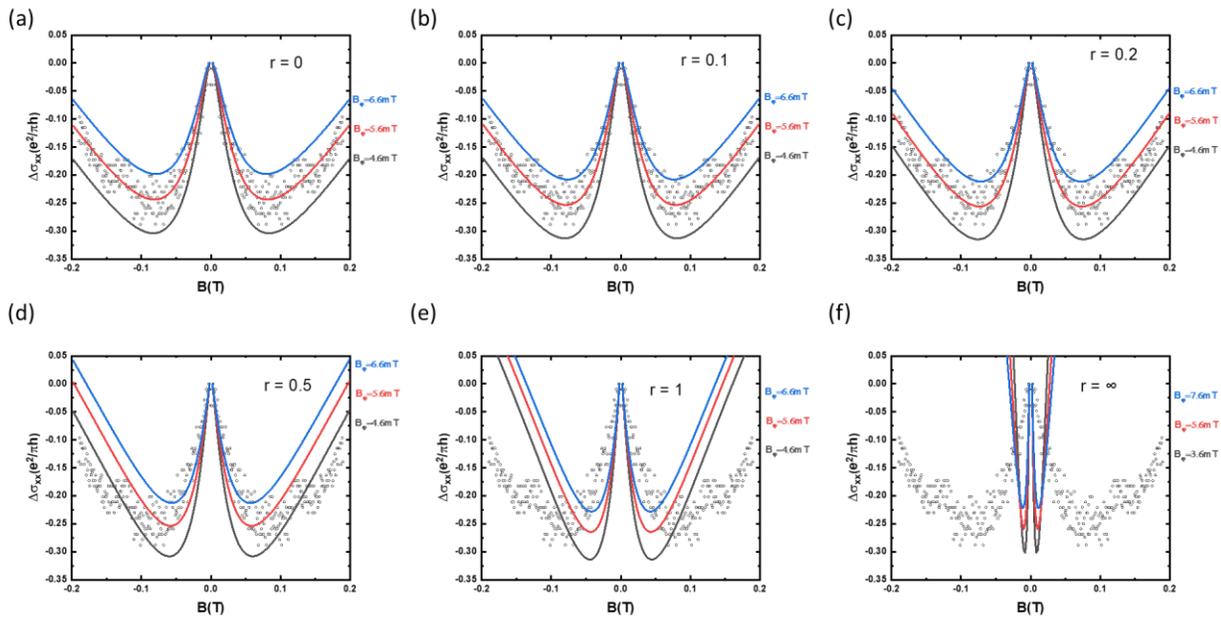

**Figure S3 (a)-(f)** The result of changing $B_\varphi$ for different $r$ values.

## Supporting Note 3: WAL in another n-type Te sample

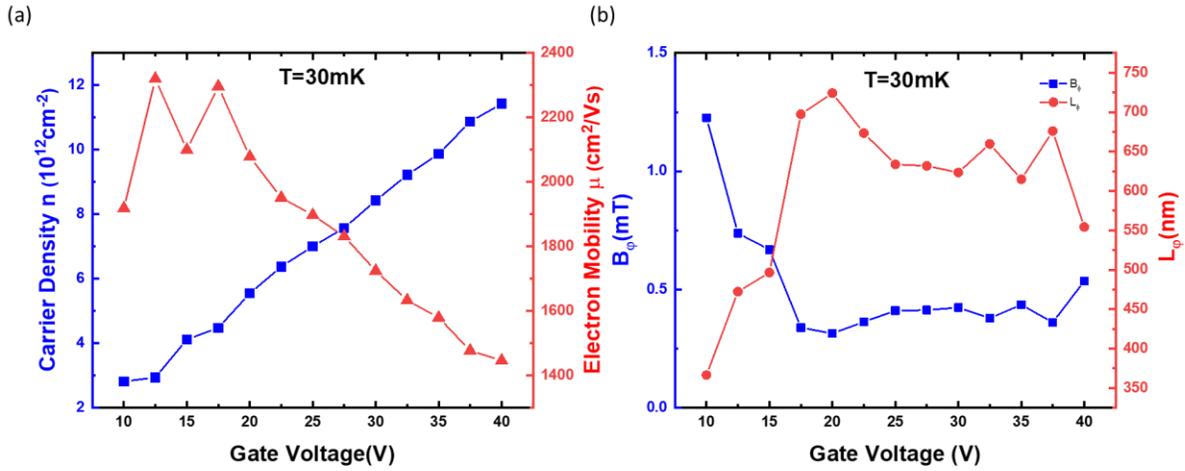

**Figure S4 (a)** Gate-dependent carrier density $n$ (blue squares, left axis) and electron mobility $\mu$ (red triangles, right axis) extracted from Hall-measurement at low temperature (30mK). **(b)** Gate-dependent $B_\varphi$ (blue squares, left axis) and phase coherence length $L_\varphi$ (red squares, right axis) extracted from WAL fitting at $T = 30mK$.

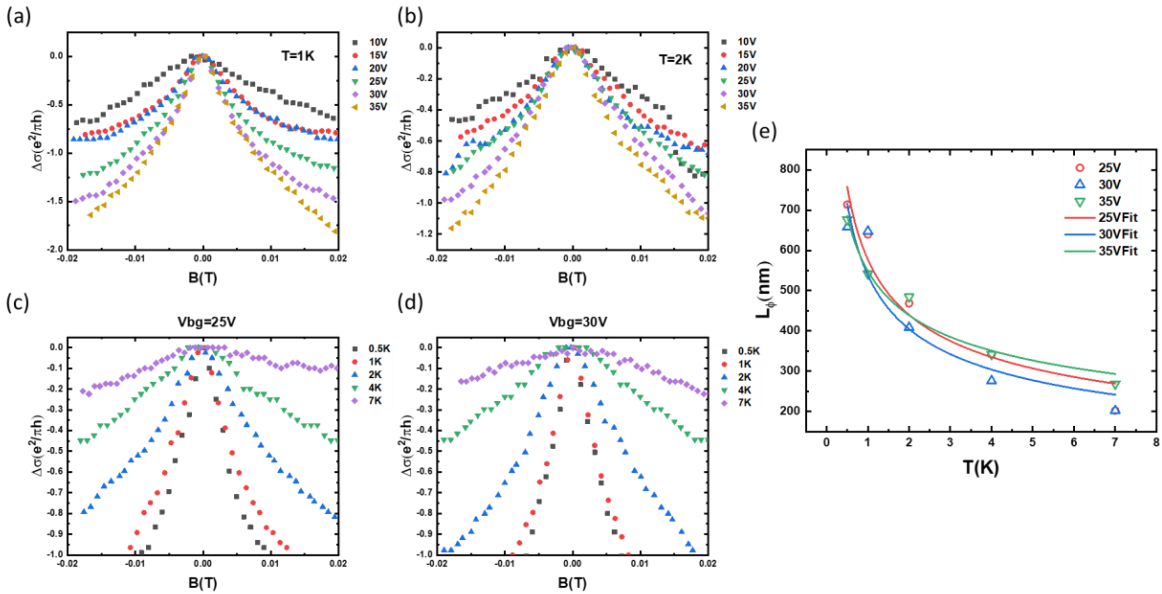

**Figure S5 (a), (b)** Gate-dependent WAL measured at $T = 1K$ and $2K$. **(c), (d)** Temperature dependent WAL measured at $V_{bg} = 25V$ and $30V$. **(e)** Phase coherence length $L_\varphi$ extracted from WAL fitting as a function of temperature.

## Supporting Note 4: WAL in P-type Te

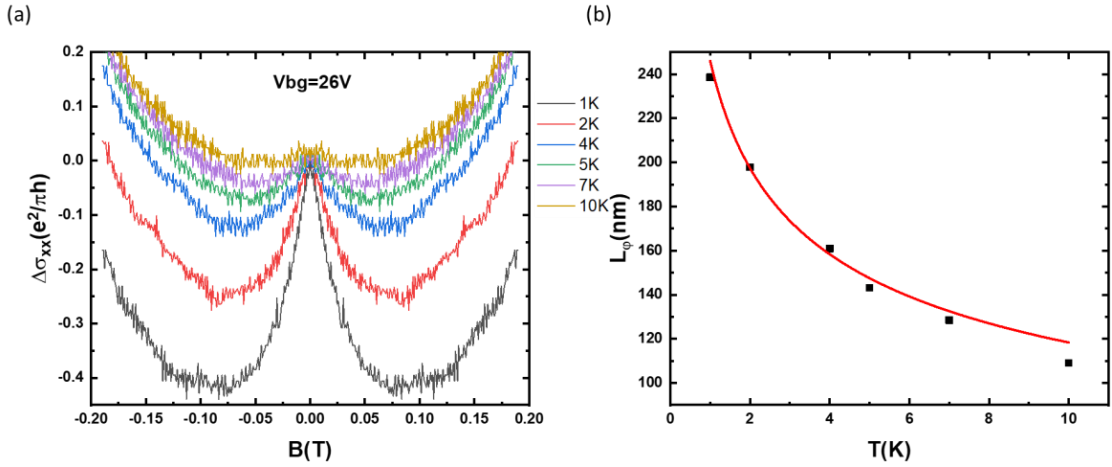

**Figure S6 (a)** Temperature dependent WAL measured at $V_{bg} = 26V$. **(b)** Phase coherence length $L_\varphi$ extracted from WAL fitting as a function of temperature.